\documentclass[aps,prd,10pt,nofootinbib,superscriptaddress,preprint]{revtex4-1}
\usepackage{amsmath,amssymb,amsfonts,dsfont,mathrsfs,amsthm,mathtools}
\usepackage{graphicx}
\usepackage{hyperref}
\usepackage{siunitx}
\usepackage{bigints}
\usepackage[english]{babel}
\hypersetup{linktocpage,colorlinks=true,urlcolor=blue,linkcolor=blue,citecolor=blue}
\usepackage{slashed}
\usepackage{float}

\newcommand*{\sign}{\operatorname{sign}}

\newcommand{\diff}[1]{\text{d}#1}

\newcommand*{\Tr}{\operatorname{Tr}}

\providecommand{\U}[1]{\protect\rule{.1in}{.1in}}

\newcommand{\be}{\begin{equation}}
\newcommand{\ee}{\end{equation}}
\newcommand{\bea}{\begin{eqnarray}}
\newcommand{\eea}{\end{eqnarray}}

\begin{document}
\title{Gravitating anisotropic merons and squashed spheres in the three-dimensional Einstein-Yang-Mills-Chern-Simons theory}

\author{Fabrizio Canfora}
\email{fabrizio.canfora@uss.cl}
\affiliation{Facultad de Ingenieria, Arquitectura y Dise\~no, Universidad San Sebastian, sede Valdivia, General Lagos 1163, Valdivia 5110693, Chile.}
\affiliation{Centro de Estudios Cient\'ificos (CECs), Casilla 1469, Valdivia, Chile}

\author{Crist\'obal Corral}
\email{crcorral@unap.cl}
\affiliation{Instituto de Ciencias Exactas y Naturales, Universidad Arturo Prat, Playa Brava 3256, 1111346, Iquique, Chile}
\affiliation{Facultad de Ciencias, Universidad Arturo Prat, Avenida Arturo Prat Chac\'on 2120, 1110939, Iquique, Chile}

\begin{abstract}
We construct the first analytic examples of self-gravitating anisotropic merons in the Einstein-Yang-Mills-Chern-Simons theory in three dimensions. The gauge field configurations have different meronic parameters along the three Maurer-Cartan $1$-forms and they are topologically nontrivial as the Chern-Simons invariant is nonzero. The corresponding backreacted metric is conformally a squashed three-sphere. The amount of squashing is related to the degree of anisotropy of the gauge field configurations that we compute explicitly in different limits of the squashing parameter. Moreover, the spectrum of the Dirac operator on this background is obtained explicitly for spin-1/2 spinors in the fundamental representation of $SU(2)$, and the genuine non-Abelian contributions to the spectrum are identified. The physical consequences of these results are discussed.

\end{abstract}

\maketitle

\section{Introduction}

The most important open problems in high-energy physics are non-perturbative in nature. For instance, when analyzing the phase diagram of QCD in the infrared~\cite{R0,R11,QGBook,R2}, perturbative methods fail and lattice QCD experiences some difficulties due to the sign problem as well as due to the magnetic field~\cite{sign1,sign3m,sign3n,sign3l,QGBook}. At low energies, a prominent role is played by the topological solitons~\cite{Manton,Shuryak,Shifman,R11}. Such classical configurations possess non-trivial topological charges, protecting them against the decay into the trivial vacuum. This is the reason why topologically nontrivial configurations in gauge theories have attracted so much attention since their discovery.

Among all the non-Abelian solitons, merons---first discovered in Ref.~\cite{merons0}---are very special. These genuine non-Abelian Yang-Mills configurations possess one-half unit of localized topological charge. Therefore, they are not observable in isolation in flat spaces since they are singular on that background. Nevertheless, merons are quite important in obtaining a correct qualitative picture of confinement; something that cannot be achieved with instantons only~\cite{merons1a,merons1b,merons1c,merons2,merons3,merons4,merons5}. Merons are also very important in the analysis of color confinement in 2+1 dimensions~\cite{merons1a,merons1b,merons1c,merons2,merons3,merons4,merons5}. Coupling the Yang-Mills theory with general relativity discloses a further remarkable effect related to merons: the gravitational
backreaction can hide the merons singularity behind the black-hole horizon making a meron-black hole system possible~\cite{merons6a,merons6b,merons6c,merons6d,Flores-Alfonso:2020ayc}.

Until now, all the available analytic meron configurations can be brought to the form $A_{\mu}=c\, U^{-1}(x)\partial_{\mu}U(x)$, where $U(x)\in SU(N)$ and $c$ is a real parameter to be fixed by solving the Yang-Mills-Einstein field equations.\footnote{The real parameter must satisfy $c\neq0,1$. Otherwise, the configuration is either trivial or pure gauge.} All the known examples are isotropic in the sense that the value of $c$ is the same for all the non-vanishing Maurer-Cartan components of
$U^{-1}\partial_{\mu}U$. To be more precise, one can expand $U^{-1}\partial_{\mu}U=\sum\Omega_{\mu}^{a}%
T^{a}$ where $T^{a}$ are the basis of the Lie algebra of $SU(N)$ and
$\Omega_{\mu}^{a}$ is the projection of $U^{-1}\partial_{\mu}U$ along the generator $T^{a}$. In principle, one could try an \emph{anisotropic meron} ansatz of the form
\begin{align}
A_{\mu}=\sum_{a=1}^{N^{2}-1}c_{(a)}\Omega_{\mu}^{a}T^{a}\,,
\end{align}
where each $\Omega_{\mu}^{a}$ has its own meronic parameter, $c_{(a)}$, with $c_{(a)}\neq c_{(b)}$ for at least one pair of indices. However, no such solution has been found. Indeed, in all the meronic configurations constructed so far, the condition $c_{(a)}=c_{(b)}$ is met and, in most of the cases, its value is fixed according to $c_{(a)}=1/2$. This issue is very important in the theory of topological solitons: given the topological charge---defined in Eq.~\eqref{ICS} below---, should one expect that the solitons with that charge are always the most symmetric given specific boundary conditions? For instance, in the $SU(2)$ Skyrme model on flat space-times~\cite{Skyrme1a,Skyrme1b,Skyrme1c,Skyrme2,Skyrme3,Skyrme4,BaMa,Manton}, it is known that the solution with baryonic charge equal to $1$ has spherical symmetry in the hedgehog sense, while the solution with baryonic charge equal to $2$ has donut shape with axial symmetry. Indeed, it is suspected that, with higher charges, only discrete symmetries are left. In the case of merons, this class of questions has not been answered yet. An interesting one---which is related to both spontaneous symmetry breaking and nontrivial topology---is the following: how anisotropic can a regular self-gravitating meron be? Indeed, as it has been emphasized, the only possibility for a meron to be regular is to couple Yang-Mills to general relativity. We will provide the answer to the above question in a very relevant situation.

In this work, we analyze the $3$-dimensional case in Euclidean signature. Unlike what happens in $2$-dimensions, in $3$-dimensions gauge theories do have local degrees of freedom. Moreover, Yang-Mills theory in 3-dimensions is simpler to analyze than its four-dimensional counterpart but, at the same time, interesting enough to disclose important nonperturbative features that are also relevant in the 4-dimensional case. Perhaps, the most important reason to focus on gauge theories in three dimensions is their relations with magnetic screening in four dimensions at high temperature \cite{R11,QGBook}. Moreover, in three dimensions, a Chern-Simons term also appears as it arises from the evaluation of the fermionic determinant~\cite{CS1,CS2,CS3,CS4a,CS4b}. Therefore, there is a lot of 4-dimensional physics in Yang-Mills-Chern-Simons theory in one dimension lower. 

Another deep motivation that makes the three-dimensional case especially interesting is the so-called~\emph{parity anomaly}~\cite{Redlich:1983kn,Redlich:1983dv,Alvarez-Gaume:1984zst,Korchemsky:1990va,Witten:2016cio,Kurkov:2017cdz,Kurkov:2018pjw}. Under homotopically nontrivial gauge transformations, the effective action of a three-dimensional $SU(N)$ gauge theory changes as the winding number in presence of an odd number of fermions species~\cite{Redlich:1983kn,Redlich:1983dv,Alvarez-Gaume:1984zst}. Nevertheless, it is possible to restore gauge invariance by means of a Pauli-Villars regularization. However, this prescription introduces a Chern-Simons term into the effective action that breaks parity invariance at the quantum level. This is known as the parity anomaly~\cite{Redlich:1983kn,Redlich:1983dv,Alvarez-Gaume:1984zst} and it emphasizes not only the important role of the Chern-Simons term but also the fact that, even without fermions, the Chern-Simons coupling cannot be any number, but it must be quantized. The remarkable physical consequences of anomalies are triggered by the presence of topologically nontrivial configurations as it happens, for instance, in four dimensions with instantons that account for the correct mass of the $\eta^{\prime}$ meson~\cite{tHooft:1976rip,tHooft:1986ooh}. Hence, due to the above reasons, we will analyze the appearance of regular anisotropic merons in Yang-Mills-Chern-Simons theory in three dimensions minimally coupled to general relativity.

There is another sound reason to analyze gravitating solitons. The AdS/CFT correspondence is a nonperturbative technique that allows one to describe both relativistic and non-relativistic strongly-coupled and correlated systems from a weakly-coupled gravity dual description~\cite{AdSCFT1,AdSCFT2,AdSCFT3}. There are many important phenomena---such as superfluidity---which, in order to be properly described with the gauge/gravity duality, need the inclusion of the backreaction of solitons~\cite{AdSCFT4,AdSCFT5}. This circumstance increases the interest in constructing gravitating topologically nontrivial solitons, keeping in mind the aforementioned issue about how symmetric gravitating solitons can be.

Among all the three-dimensional geometries that have been considered for applications of the AdS/CFT correspondence, squashed spheres have played a very important role. The topology is, of course, still that of $\mathbb{S}^3$. Nevertheless, these geometries possess (at least) one deformation parameter which makes such geometries to be neither maximally symmetric nor of constant curvature. Hence, there are many possible AdS/CFT settings where such a deformation parameter encodes relevant information on the dual field theory allowing, in some cases, exact checks of the gauge/gravity correspondence without conformal invariance. For instance, on a squashed three-sphere, one can show explicitly that the expansion of the free energy in the squashing parameter is related to the stress-energy tensor three-point function. Indeed, one can compute the correlators of Higgs and Coulomb branch operators of $\mathcal{N} = 4$ supersymmetric theories and so on~\cite{squashed1a,squashed1b,squashed2,squashed3,squashed4,squashed5a,squashed5b,squashed6a,squashed6b,squashed6.1,squashed6.2a,squashed6.2b,squashed6.2c}. In general, all these powerful results, which heavily employ the localization techniques introduced in Ref.~\cite{squashed7}, show the ability of the squashed sphere geometry to describe symmetry breaking patterns, which is one of the main topics of interest as far as the present paper is concerned.

Due to the fact that in most cases the squashed sphere is considered a fixed background on which the field theory of interest is studied, there is a very important question that is usually not analyzed in detail in the literature: are there reasonable theories that admit the three-dimensional squashed sphere as a solution? If the answer to this question is negative, then the aforementioned references would partially lose their strength. On the other hand, if the answer is affirmative, then not only the known results on field theories on squashed spheres would acquire an even sounder physical basis, but also one would be able to derive the dependence of the squashing parameter in terms of other relevant parameters of the problems of interest; this cannot be done if the squashed sphere is treated as a fixed background. Specifically, we will be able to derive the exact dependence of the squashing parameter on the cosmological constant, on the volume of the squashed sphere, as well as on the other parameters, so that the expansion on the squashing parameter acquires a very precise meaning in terms of the coupling constants and parameters of the problem. 

In this paper, we will show that the answer is, indeed, affirmative: squashed spheres emerge as topologically nontrivial solutions of the Einstein-Yang-Mills-Chern-Simons theory in three dimensions. In particular, we will show that anisotropic merons are natural sources of the latter and
that the amount of squashing is closely related to the amount of anisotropy of these non-Abelian configurations. Moreover, the meronic nature of the Yang-Mills fields allows us to determine exactly the spectrum of the Dirac operator of the spin-1/2 spinors which are coupled both to the gravitational and non-Abelian gauge fields. One of the nice features of the latter is that it allows one to single out explicitly the genuine non-Abelian contributions when comparing it to the analogous spectrum for a spin 1/2 field with an Abelian gauge field.

The paper is organized as follows. In Sec.~\ref{sec:isomeron}, we review the $SU(2)$ isotropic meron in Yang-Mills theory on a four-dimensional Euclidean space. In Sec.~\ref{sec:selfgravmeron}, we present the self-gravitating anisotropic meron in the three-dimensional Einstein-Yang-Mills-Chern-Simons theory and demonstrate that their backreaction supports a conformally squashed three-sphere as a solution. Section~\ref{sec:action} is devoted to show the global properties of the solution, such as the action and topological invariants. In Sec.~\ref{sec:spectrum}, we compute the spectrum of the Dirac operator and provide an analytic expression for their eigenvalues. Finally, we present our conclusions and perspectives in Sec.~\ref{sec:conclusions}. Appendix~\ref{sec:eta} is included for discussing the implications of these results on the Atiyah-Patodi-Singer (APS) $\eta$-invariant~\cite{APS} that measures the spectral asymmetry generated by these non-Abelian anisotropic configurations.

\section{Isotropic meron in SU(2) Yang-Mills theory\label{sec:isomeron}}

The isotropic meron solution of $SU(2)$ Yang-Mills theory on a
$\mathbb{R}^{4}$ background was first found in Ref.~\cite{merons0}. Here, we review the latter for the sake of completeness and fix conventions. To do so, we define the gauge connection as $A_{\mu}=A_{\mu}^{i}t_{i}$ where $t_{i}=-\frac{i}{2}%
\tau_{i}$ are the generators of $SU(2)$ with $\tau_{i}$ being the Pauli
matrices. The latter satisfies the Lie algebra and the anticommutation
relation
\begin{align}
    \lbrack\tau_{i},\tau_{j}]=2i\epsilon_{ijk}\tau^{k}%
\,\;\;\;\;\;\mbox{and}\;\;\;\;\;\{\tau_{i},\tau_{j}\}=2\delta_{ij}%
\,\mathbb{I}\,,
\end{align}
respectively, where $\delta_{ij}$ and $\epsilon_{ijk}$ are the two invariant tensors of $SU(2)$; these relations imply $\tau_{i}\tau_{j}=\delta_{ij}\mathbb{I}%
+i\epsilon_{ijk}\tau^{k}$. From these conditions one can show that
$\Tr(t_{i}t_{j})=-\frac{1}{2}\delta_{ij}$ and $\Tr(t_{i}t_{j}t_{k})=-\tfrac
{1}{4}\epsilon_{ijk}$ are satisfied. Additionally, the non-Abelian field strength for the Yang-Mills fields is defined by
\begin{align}\label{Fmunu}
F_{\mu\nu}=\partial_{\mu}A_{\nu} - \partial_{\nu}A_{\mu}+ \left[  A_{\mu
},A_{\nu}\right]  \,.
\end{align}
The dynamics of these non-Abelian fields is dictated by the Yang-Mills action principle that, on a $D$-dimensional Riemannian manifold $(\mathcal{M},g_{\mu\nu})$, can
be written as
\begin{align}
\label{actionYM}I_{\mathrm{YM}} = \frac{1}{2e^{2}} \int_{\mathcal{M}}%
\text{d}^{D}x\sqrt{|g|}\,\Tr\left(  F_{\mu\nu}F^{\mu\nu}\right)  \,,
\end{align}
where $e$ is the $SU(2)$ coupling constant and $g=\det g_{\mu\nu}$ is the metric determinant. This is a nonlinear functional of the gauge fields that depends only on first-order derivatives thereof. The field equations are obtained by performing arbitrary variations of the action~\eqref{actionYM} with respect to the gauge connection, $A_{\mu}$, giving
\begin{align}
\label{eomA}\mathcal{E}^{\mu}  &  \equiv\nabla_{\mu}F^{\mu\nu} + \left[
A_{\mu},F^{\mu\nu} \right]  = 0\,.
\end{align}

In order to show why the value of $c=1/2$ is indeed special for the meronic configurations, let us discuss the simplest example of meron in Euclidean four-dimensional spacetimes. In order to do so, it is convenient to define the Maurer-Cartan left-invariant $1$-forms of $SU(2)$, $\sigma_{i}$, as
\begin{subequations}
\label{MCformsu2}%
\begin{align}
\sigma_{1} &  =\cos\psi\,\text{d}\vartheta+\sin\vartheta\,\sin\psi
\,\text{d}\varphi\,,\\
\sigma_{2} &  =-\sin\psi\,\text{d}\vartheta+\sin\vartheta\,\cos\psi
\,\text{d}\varphi\,,\\
\sigma_{3} &  =\text{d}\psi+\cos\vartheta\,\text{d}\varphi\,.
\end{align}
\end{subequations}
They satisfy the relation $\diff{}\sigma_{i}+\frac{1}{2}\epsilon_{ijk}\sigma^{j}\wedge\sigma^{k}=0$. The line element of $\mathbb{R}^{4}$ can be written in terms of these $1$-form as
\begin{align} 
   \diff{s^2} &= \diff{r^2} + \frac{r^2}{4}\,\left(\sigma_1^2+\sigma_2^2+\sigma_3^2\right) \,, \label{R4}
\end{align}
where $0\leq r<\infty$, $0\leq\vartheta\leq\pi$, $0\leq\varphi\leq2\pi$, and
$0\leq\psi\leq4\pi$. 

To solve the Yang-Mills equations on this background, we consider the ansatz\footnote{In the language of differential forms, the non-Abelian field strength in Eq.~\eqref{Fmunu} can be seen as the components of the $2$-form $F=\tfrac{1}{2}F^i_{\mu\nu}t_i\,\diff{x^\mu}\wedge\diff{x^\nu} =\diff{A}+\frac{1}{2}[A,A]$.}
\begin{align}\label{meron}
A = A_\mu^i t_i\,\diff{x^\mu} = \sum_{i=1}^3 c_{(i)}\,\sigma^i\,t_i\,.
\end{align}
In principle, the three real parameters $c_{(i)}$ (with $i=1,2,3$) could be different. This implies that $c_{(i)}\neq c_{(j)}$ for at least one pair of indices $(i\neq j)$. However, the known meronic solution to the Yang-Mills equations~\eqref{eomA} on flat space~\eqref{R4}, as well as the self-gravitating merons in Ref.~\cite{merons6a,merons6b,merons6c,merons6d} are isotropic, namely,
\begin{align}\label{csolisomeron}
c_{(i)}= \frac{1}{2}\,,\;\;\;\;\; \forall i=1,2,3\ .    
\end{align}
This configuration is neither trivial nor pure gauge as it can be seen by checking that the non-Abelian field strength $2$-form is given by
\begin{align}\label{Fisomeronic}
    F = -\frac{1}{8}\,\epsilon^{ijk}\,\sigma_i\wedge\sigma_j\,t_k\,,
\end{align}
where $\wedge$ is the wedge product of differential forms. Then, the isotropic meron configuration found in Ref.~\cite{merons0} represents a solution to the Yang-Mills equations~\eqref{eomA} with constant field strength.

The topological charge of this configuration is characterized by the Chern-Simons invariant for the $SU(2)$ group on the codimension-1 hypersurfaces of constant $r$, that is,
\begin{align}\label{ICS}
    I_{\rm CS} = \frac{1}{8\pi^2} \int_{\mathbb{S}^3}\Tr\left(A\wedge\diff{A}+\frac{2}{3}A\wedge A\wedge A\right).
\end{align}
In the case of the
meronic solution~\eqref{meron} with the isotropic values of the constants given in Eq.~\eqref{csolisomeron}, one finds that the Chern-Simons invariant is
\begin{align}\label{ICSisomeron}
I_{\rm CS}=\frac{1}{2}\,.
\end{align}
Therefore, since this configuration possesses a nontrivial topology measured by a fractional topological charge, it cannot be deformed continuously to the topologically trivial vacuum. Moreover, it is believed that instantons can be thought of as a bound state of two merons~\cite{merons1a}.

\section{Gravitating anisotropic meron on squashed spheres\label{sec:selfgravmeron}}

In contrast to the isotropic meron, its anisotropic counterpart does not exist in flat space. Motivated by the interest in gravitating solitons as discussed in the introduction, we look for self-gravitating anisotropic merons in Einstein-Yang-Mills theory whose dynamics in arbitrary dimensions is dictated by the action principle
\begin{align}\label{action}
 I_{\rm EYM} &= \kappa\int_{\mathcal{M}}\diff{^Dx}\sqrt{|g|}\,\left(R-2\Lambda \right) + I_{\rm YM} \,,
\end{align}
where $\kappa=(16\pi G)^{-1}$ denotes the gravitational constant, $\Lambda$ represents the cosmological constant, $g=\det g_{\mu\nu}$ is the metric determinant, and $R=g^{\mu\nu}R_{\ \mu\lambda\nu}^{\lambda}$ is the Ricci scalar.

The field equations are obtained by performing stationary variations with respect to the metric and the Yang-Mills field, giving
\begin{align}
\label{eomg}\mathcal{E}_{\mu\nu}  &  \equiv R_{\mu\nu} - \frac{1}{2}g_{\mu\nu
}R + \Lambda g_{\mu\nu} - \frac{1}{2\kappa}T_{\mu\nu} = 0\,,
\end{align}
and Eq.~\eqref{eomA}, respectively, where
\begin{align}
\label{Tmunu}T_{\mu\nu}  &  = -\frac{2}{e^{2}}\Tr\left(  F_{\mu\lambda}F_{\nu
}^{\ \lambda} - \frac{1}{4}g_{\mu\nu}F_{\lambda\rho}F^{\lambda\rho} \right)
\,,
\end{align}
is the stress-energy tensor of the Yang-Mills fields.

In three dimensions, there is an additional term that can be added to the Einstein-Yang-Mills action, which is called the Chern-Simons invariant given in Eq.~\eqref{ICS}. If we consider the action principle of such a theory, i.e.
\begin{align}\label{actionEYMCS}
I=I_{\mathrm{EYM}}+\frac{4\pi^{2}k}{e^{2}}I_{\mathrm{CS}}\ ,    
\end{align}
with $k$ being the Chern-Simons coupling constant the latter term modifies the Yang-Mills equation according to
\begin{align}
    \label{eomACS}
 \tilde{\mathcal{E}}^\mu &\equiv \mathcal{E}^\mu + k\,\varepsilon^{\mu\nu\lambda}F_{\nu\lambda} = 0\,,
\end{align}
where $\mathcal{E}^{\mu}$ is defined in Eq.~\eqref{eomA} and $\varepsilon^{\mu\nu\lambda}$ is the alternating Levi-Civita tensor in three dimensions. The stress-energy tensor in Eq.~\eqref{Tmunu}, however, is not modified due to the topological nature of the Chern-Simons term in Eq.~\eqref{ICS}: it is independent of the metric.

\begin{figure}[H]
    \centering
    \includegraphics[scale=0.4]{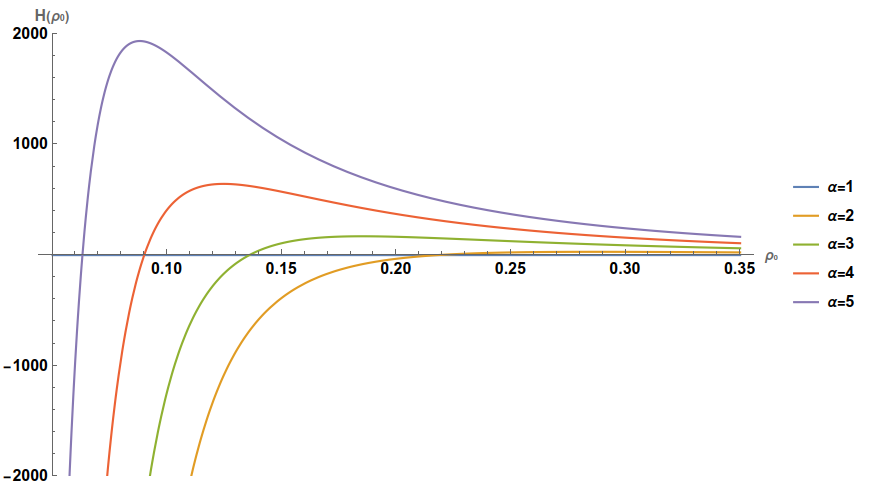}
    \caption{Polynomial equation of degree
5 for $\rho_{0}$ in terms of $\alpha$, $\kappa$, $e^{2}$, and $k$, say,
$H(\rho_{0})=H(\rho_{0};\alpha,\kappa,e^{2},k)=0$. In this plot, we have fixed
$\kappa=(16\pi)^{-1}$, $e=1$, and $k=1$.}
    \label{fig:Hrho}
\end{figure}

As explained in detail in the Introduction, we are interested in
squashed spheres. Thus, the natural ansatz for the metric is a conformally squashed sphere given by 
\begin{align}\label{metricansatz}
    \diff{s^2} = \rho_0^2\left(\sigma_1^2 + \sigma_2^2 +\alpha^2\sigma_3^2  \right)\,,
\end{align}
where $\rho_0$ is a constant conformal factor related to the volume of this three-dimensional space and $\alpha$ is the squashing parameter. This line element, alongside the anisotropic meron ansatz in Eq.~\eqref{meron}, implies that the Yang-Mills-Chern-Simons field equations~\eqref{eomACS} reduce to three linearly independent algebraic equations for the parameters $c_{(i)}$; they are
\begin{subequations}
\label{YMcomp}%
\begin{align}
\alpha^{2}c_{(2)}\left(  c_{(1)}c_{(2)}-c_{(3)}\right)  -2k\alpha\rho
_{0}\left(  c_{(2)}c_{(3)}-c_{(1)}\right)  +c_{(1)}c_{(3)}^{2}-2c_{(2)}%
c_{(3)}+c_{(1)} &  =0\,,\\
\alpha^{2}c_{(1)}\left(  c_{(1)}c_{(2)}-c_{(3)}\right)  -2k\alpha\rho
_{0}\left(  c_{(1)}c_{(3)}-c_{(2)}\right)  +c_{(2)}c_{(3)}^{2}-2c_{(1)}%
c_{(3)}+c_{(2)} &  =0\,,\\
\alpha^{2}\left(  c_{(1)}c_{(2)}-c_{(3)}\right)  +2k\alpha\rho_{0}\left(
c_{(1)}c_{(2)}-c_{(3)}\right)  -c_{(1)}^{2}c_{(3)}-c_{(2)}^{2}c_{(3)}%
+2c_{(1)}c_{(2)} &  =0\,.
\end{align}
\end{subequations}
This system distinguishes between two different solutions: (i) $c_{(2)}\neq c_{(1)}$ and (ii) $c_{(2)}=c_{(1)}$. Nevertheless, the off-diagonal components of the Einstein-Yang-Mills-Chern-Simons equations~\eqref{eomg} demand that only the second case is compatible with a self-gravitating anisotropic meron. Then, focusing on that case, we find
that the solution is given by
\begin{align}\label{YMsol}
     c_{(1)} &= c_{(2)}\,, &
    c_{(2)}^2 &= \frac{1}{16}\left(\Xi \pm\sqrt{\Delta}\right)\,, &
    c_{(3)} &= \frac{4c_{(2)}^2\left(\alpha\Phi+4 \right)}{\Xi+4\alpha\Phi\pm\Phi\sqrt{\Delta}}\,,
\end{align}
where we have defined
\begin{align}
\Xi&=8k^2\rho_0^2 - 8k\alpha\rho_0-6\alpha^2    \,,\\
\Delta&=4k^2\rho_0^2+20k\alpha\rho_0+9\alpha^2+16\,,\\
\Phi &= 4k\rho_0+2\alpha\,.
\end{align}
The Einstein-Yang-Mills-Chern-Simons equations~\eqref{eomg}, in turn, lead to the following set of linearly independent algebraic equations
\begin{subequations}\label{eomgalgebraic}
\begin{align}
\label{eomg1}
\Lambda &  =\frac{4-3\alpha^{2}}{4\rho_{0}^{2}}+\frac{2c_{(2)}^{2}\left(
c_{(3)}-1\right)  ^{2}-\alpha^{2}\left(  c_{(2)}^{2}-c_{(3)}\right)  }{4\kappa
e^{2}\alpha^{2}\rho_{0}^{4}}\,,\\
\label{eomg2}
\Lambda &  =\frac{\left(  c_{(2)}^{2}-c_{(3)}\right)  ^{2}}{4\kappa e^{2}%
\rho_{0}^{4}}+\frac{\alpha^{2}}{4\rho_{0}^{2}}\,.
\end{align}
\end{subequations}
In principle, these equations fix the squashing parameter $\alpha$ as a function of the cosmological constant and the other parameters. However, it is much easier to express the cosmological constant $\Lambda$ in terms of the squashing parameter $\alpha$ rather than the other way around. Then, replacing Eq.~\eqref{eomg2} into Eq.~\eqref{eomg1} alongside the particular values for the Yang-Mills fields found in Eq.~\eqref{YMsol}, we find an algebraic master equation of degree 5 for $\rho_{0}$ in terms of $\alpha$, $\kappa$, $e^{2}$, and $k$, say, $H(\rho_{0};\alpha,\kappa,e^{2},k)=0$, with
\begin{align}
H(\rho_{0};\alpha,\kappa,e^{2},k)  & =p_{5}\rho_{0}^{5}+p_{4}\rho_{0}%
^{4}+p_{3}\rho_{0}^{3}+p_{2}\rho_{0}^{2}+p_{1}\rho_{0}+p_{0}\, ,
\end{align}
where we have defined
\begin{subequations}
\begin{align}
p_{5}  & =16k^3\alpha\left(2e^2\kappa-k^2\right)\,, \\
p_{4}  & = \left(8e^4\kappa^2+124e^2k^2\kappa-16k^4\right)\alpha^2-4\left(2e^2\kappa-k^2\right)\left(e^2\kappa-2k^2\right)\,,\\
p_{3}  & = 4k\left(27e^2\kappa-k^2\right)\alpha^3+4k\left(21e^2\kappa+2k^2\right)\alpha\,, \\
p_{2}  & = 27\alpha^4e^2\kappa +\left(45e^2\kappa+6k^2\right)\alpha^2 + 8\kappa e^2 + 8k^2 \,, \\
p_{1}  & = 0 \,, \\
p_{0}  & = -2 .
\end{align}
\end{subequations}
It is remarkable that one can reduce the full set of Einstein-Yang-Mills-Chern-Simons field equations in a topologically nontrivial sector to just one algebraic master equation. Since the latter is a polynomial of order 5 in $\rho_0$ in the generic case, it can be solved only numerically as shown in Figure~\ref{fig:Hrho}.
Notice that the larger is the squashing, the smaller is $\rho_{0}$; the latter represents the volume of the space. On the other hand, we find that the cosmological constant is a monotonically increasing function of the squashing parameter. For instance, the behavior of the cosmological constant as a function of the squashing parameter reveals that
\begin{align}\label{Lalpha}
   \Lambda =  \begin{cases}
       \kappa e^2 + 4e^2\alpha\kappa + \frac{15}{2}\alpha^2 e^2\kappa + \mathcal{O}(\alpha^3) & \mbox{as} \;\;\; \alpha\to0\,, \\
       35\kappa e^2+\frac{544}{5}\kappa e^2(\alpha-1)+\mathcal{O}\left((\alpha-1)^2\right) & \mbox{as} \;\;\; \alpha\to1\,, \\
       \frac{27}{8}\alpha^6 e^2\kappa+\frac{117}{8}\alpha^4 e^2 \kappa + 15\alpha^2 e^2\kappa + \frac{53}{27}\kappa e^2 + \mathcal{O}(\alpha^{-2}) & \mbox{as} \;\;\;  \alpha\to\infty\,.
   \end{cases}
\end{align}   
Thus, we conclude that the cosmological constant is positive definite for a squashing parameter in the range $0<\alpha<\infty$.

When the Chern-Simons term is absent, namely $k=0$, the master equation can be solved analytically.\footnote{This is similar to what happens in Refs.~\cite{merons6b,Canfora:2021ghv}, where the solution is smoothly connected to $k=0$.} Then, one arrives at the following
solution
\begin{subequations}
\label{solk0}%
\begin{align}
c_{(1)} &  =c_{(2)}\,,\\
c_{(2)} &  =\frac{1}{4}\left[  2\alpha\left(  \sqrt{9\alpha^{2}+16}%
-3\alpha\right)  \right]  ^{1/2}\,,\\
c_{(3)} &  =\frac{\left(  \alpha^{2}+2\right)  \left(  \sqrt{9\alpha^{2}%
+16}-3\alpha\right)  }{2\left(  \sqrt{9\alpha^{2}+16}+\alpha\right)  }\,,\\
\rho_{0}^{2} &  =\frac{27\alpha^{6}+117\alpha^{4}+144\alpha^{2}+32-\left(
9\alpha^{5}+31\alpha^{3}+24\alpha\right)  \sqrt{9\alpha^{2}+16}}{8\kappa
e^{2}\left[  8-3\alpha^{2}-5\alpha^{4}-\alpha\left(  \alpha^{2}-1\right)
\sqrt{9\alpha^{2}+16}\right]  }\,,
\end{align}
\end{subequations}
and $\Lambda$ is fixed in terms of $\alpha$ as one can see from Eq.~\eqref{eomgalgebraic}. From these expressions, one can check that the metric function behaves as $\rho_{0}=\frac{1}{2e}\sqrt{\frac{2}{\kappa}}+\mathcal{O}(\alpha)$ in the small-squashing limit. In the large-squashing limit, on the other hand, its behavior is $\rho_{0}\sim\mathcal{O}(\alpha^{-2})$. Moreover, for large squashing, the amount of
anisotropy of the merons is fixed, while for $\alpha\to0$ the anisotropy is very large. This can be seen by analyzing the behavior of the ratio 
\begin{align}
\frac{c_{(3)}}{c_{(1)}} = \frac{\left(\alpha^2+2 \right)\left[2\left(\sqrt{9\alpha^2+16} - 3\alpha\right) \right]^{1/2}}{\sqrt{\alpha}\left(\sqrt{9\alpha^2+16}+\alpha \right)}  \,,
\end{align}
for different values of the squashing parameter. Indeed, it can be easily check that its value is equal to 1 when $\alpha\to1$, recovering the isotropic meron of Sec.~\ref{sec:isomeron}. Remarkably, when the three-sphere is very squashed, the degree of anisotropy is fixed according to
\begin{align}\label{anisotropyalphagrande}    \lim_{\alpha\to\infty}\frac{c_{(3)}}{c_{(1)}} = \frac{1}{\sqrt{3}}\,.
\end{align}
In the small-squashing limit, however, the merons become very anisotropic, since the amount of anisotropy becomes
\begin{align}
    \frac{c_{(3)}}{c_{(1)}} = \sqrt{\frac{2}{\alpha}} - \frac{5\sqrt{2}}{8}\alpha^{1/2} + \mathcal{O}(\alpha^{3/2})\;\;\;\;\; \mbox{as} \;\;\;\;\; \alpha\to0.
\end{align}


To the best of our knowledge, this solution represents the first self-gravitating anisotropic meron in Einstein-Yang-Mills-Chern-Simons theory and it is completely determined by the value of the squashing parameter $\alpha$. In other words, since we could have expressed $\alpha$ in terms of the cosmological constant (but we have chosen the other way around as the corresponding algebraic expressions are simpler), the conclusion is that the present gravitating anisotropic regular meron has no integration constants. The three main parameters characterizing the solution, i.e. $\rho_{0}$, the squashing parameter, and the amount of anisotropy are fixed in terms of the
coupling constants of the theory. This is to be expected as gravitating solitons usually possess fewer integration constants than, for instance, black holes, where almost always the mass is an integration constant. 

\section{Action and topological terms\label{sec:action}}

In this section, we analyze the global properties of the analytical anisotropic meron solution in Eq.~\eqref{solk0}. First, notice that the Euclidean on-shell action~\eqref{actionEYMCS} becomes that of Eq.~\eqref{action} when $k=0$. Then, evaluating the latter on the self-gravitating anisotropic meron solution we find
\begin{align}
I_{E}=\frac{8\pi^{2}\left[  \kappa e^{2}\rho_{0}^{2}\alpha^{4}+\left(  4\kappa
e^{2}\rho_{0}^{2}\left[  \Lambda\rho_{0}^{2}-1\right]  +\left[  c_{(2)}%
^{2}-c_{(3)}\right]  ^{2}\right)  \alpha^{2}+2c_{(2)}^{2}\left(
c_{(3)}-1\right)  ^{2}\right]  }{e^{2}\alpha\rho_{0}}\,,
\end{align}
where the values of Eq.~\eqref{solk0} hold on-shell. It can be checked that
$-I_{E}$ is bounded from below, whose global minimum is achieved when
$\alpha=0$. Moreover, the latter is well defined in the limit $\alpha
\rightarrow1$, since
\[
\lim_{\alpha\rightarrow1}I_{E}^{2}=\frac{32\,\pi^{3}}{5\,e^{2}\,G}\,.
\]
In this case, the meron becomes isotropic since $c_{(i)}=1/2$ for $i=1,2,3$.
On the other hand, the Chern-Simons term~\eqref{ICS} upon evaluation on the
anisotropic meron gives
\begin{align}\notag
I_{\mathrm{CS}}&=c_{(1)}^{2}+c_{(2)}^{2}+c_{(3)}^{2}-2c_{(1)}c_{(2)}c_{(3)} \\ 
&= \frac{9\alpha^6+38\alpha^4+50\alpha^2+16-(3\alpha^5+10\alpha^3+6\alpha)\sqrt{9\alpha^2+16}}{10\alpha^2+16 + 2\alpha\sqrt{9\alpha^2+16}} \, .\label{ICSanisomeron}%
\end{align}
Thus, one can see that the value of this topological term depends solely on the Yang-Mills fields through $c_{(i)}$'s. As one would expect, the above expression does not depend \emph{explicitly} on the metric being the Chern-Simons term a topological invariant. However, since the gauge field configuration does depend explicitly on the metric, due to the coupled Einstein-Yang-Mills-Chern-Simons field equations, it turns out that $I_{\mathrm{CS}}$ depends on $\alpha$, as one can verify directly from Eq.~\eqref{ICSanisomeron}. On the other hand, this should not be surprising as changing $\alpha$ means changing the theory as $\alpha$ is uniquely fixed in terms of the coupling constants of the theory. In the limit $\alpha\rightarrow1$ one can check that its value matches precisely the one in Eq.~\eqref{ICSisomeron}, which is $I_{\mathrm{CS}}=1/2$.
Remarkably, one can check that its value is finite as $\alpha\rightarrow0$ and
$\alpha\rightarrow\infty$, giving $I_{\mathrm{CS}}=1$ and $I_{\mathrm{CS}%
}=5/9$, respectively. Therefore, we conclude that the Chern-Simons form
in Eq.~\eqref{ICSanisomeron} is bounded from below for $0<\alpha<\infty$. 

The gravitational Chern-Simons term, on the other hand, is defined as
\begin{align}
    I_{\rm GCS} = \frac{1}{8\pi^2}\int\Tr\left(\omega\wedge\diff{\omega} + \frac{2}{3}\omega\wedge\omega\wedge\omega\right)\,,
\end{align}
where $\omega^{ab}=\omega^{ab}{}_{\mu}\text{d}x^{\mu}$ is the Levi-Civita connection $1$-form satisfying the torsion-free condition $\text{d}%
e^{a}+\omega_{\ b}^{a}\wedge e^{b}=0$ and $e^{a}=e_{\mu}^{a}\text{d}x^{\mu}$
denotes the dreibein $1$-form related to the metric in Euclidean signature
through $g_{\mu\nu}=\delta_{ab}e_{\mu}^{a}e_{\nu}^{b}$. For the metric ansatz in Eq.~\eqref{metricansatz}, we see that the gravitational Chern-Simons form does not depend on $\rho_{0}$ and it is explicitly given in terms of the squashing parameter as
\begin{align}
    I_{\rm GCS} = 2\alpha^4 - 4\alpha^2 + 4\,.
\end{align}
This functional possesses a global minimum at $\alpha=1$ and it is finite as
$\alpha\rightarrow0$. However, it diverges as $\alpha\rightarrow\infty$. In the latter limit, however, the geometry becomes singular, even though the amount of anisotropy remains fixed as shown in  Eq.~\eqref{anisotropyalphagrande}. In conclusion, for both the non-Abelian and the gravitational Chern-Simons invariants, the result depends on the squashing. Nevertheless, it is known that a topological invariant is composed by the sum of the Chern-Simons forms and the APS $\eta$-invariant (see for instance Ref.~\cite{Bakas:2011nq}). In the next section, we obtain the spectrum of the Dirac operator that will be the necessary starting point to compute the latter.

\section{Eigenvalues of the Dirac operator on the anisotropic meron background\label{sec:spectrum}}

The spectrum of the Dirac operator on the background of the gravitating anisotropic meron can be determined analytically by algebraic methods. In this case, the Dirac operator in the fundamental representation of $SU(2)$ is
\begin{align}
    \slashed{D} = \gamma^a E_a^\mu \left(\partial_\mu + \frac{1}{4}\omega^{bc}{}_\mu\,\gamma_{[b}\gamma_{c]} + iA_\mu \right)\,,
\end{align}
where Latin characters denote $SO(3)$ internal indices and $\gamma^{a}$ are the Dirac $\gamma$-matrices satisfying the Clifford algebra in Euclidean signature, i.e. $\{\gamma_{a},\gamma_{b}\}=2\delta_{ab}\mathbb{I}$. The vielbein $1$-form is defined as $e^{a}=e_{\mu}^{a}\text{d}x^{\mu}$ whose components are related with the metric through $g_{\mu\nu}=\delta_{ab}e_{\mu}^{a}e_{\nu}^{b}$, $E_{a}=E_{a}^{\mu}\partial_{\mu}$ is a dual vector to the vielbein such that $\langle e^{a},E_{b}\rangle=\delta_{b}^{a}$, the spin connection $1$-form is defined as $\omega^{ab}=\omega^{ab}{}_{\mu}\text{d}x^{\mu}$ and it satisfies the torsion-free condition $\diff{}e^{a}+\omega^{a}{}_{b}\wedge e^{b}=0$, and the non-Abelian gauge field $1$-form $A_\mu$ is defined in Eq.~\eqref{meron}. The main reason behind the fact that the spectrum can be solved by algebraic methods without the need to solve differential equations is related to the fact that the meronic configurations can be expressed in terms of the Maurer-Cartan forms which are the same used to construct the dreibein. Moreover, the present approach to determine the spectrum of the Dirac operator also allows us to identify the genuine non-Abelian contributions to the eigenvalues.

To evaluate the Dirac operator on the three-dimensional background of Sec.~\ref{sec:selfgravmeron}, first, we notice that the Dirac operator transforms homogeneously under conformal transformations. Thus, for
computing its spectrum, we can set $\rho_{0}=1$ in Eq.~\eqref{metricansatz} without loss of generality. Then, we can choose a dreibein basis as
\begin{align}
    e^1 &= \sigma_1\,, & e^2 &= \sigma_2\,, & e^3 &= \alpha\,\sigma_3\,.
\end{align}
The vector basis dual to the dreibein $1$-form can be defined as
\begin{align}
    E_1 &= \Sigma_1\,, & E_2 &= \Sigma_2\,, & E_3 &= \alpha^{-1}\Sigma_3\,,
\end{align}
where $\Sigma_{i}$ are the dual to the left-invariants forms of $SU(2)$, i.e.
$\langle\sigma^{i},\Sigma_{j}\rangle=\delta_{j}^{i}$, which, using the parametrization given in Eq.~\eqref{MCformsu2}, they are explicitly given by
\begin{subequations}
\begin{align}
\Sigma_{1} &  =-\cot\vartheta\sin\psi\,\partial_{\psi}+\cos\psi\,\partial
_{\vartheta}+\frac{\sin\psi}{\sin\vartheta}\,\partial_{\varphi}\,,\\
\Sigma_{2} &  =-\cot\vartheta\cos\psi\,\partial_{\psi}-\sin\psi\,\partial
_{\vartheta}+\frac{\cos\psi}{\sin\vartheta}\partial_{\varphi}\,,\\
\Sigma_{3} &  =\partial_{\psi}\,.
\end{align}
\end{subequations}
To obtain the components of the spin connection, we solve the first-order torsion-free condition $\diff{}e^{a}+\omega^{a}{}_{b}\wedge e^{b}=0$ and find
\begin{align}
    \omega^{12} &= \frac{\alpha^2-2}{2}\,\sigma_3\,, & \omega^{13} &= \frac{\alpha}{2}\,\sigma_2\,, & \omega^{23} &= - \frac{\alpha}{2}\,\sigma_1\,.
\end{align}
The Dirac $\gamma$-matrices in three dimensions are the Pauli matrices, i.e.
$\gamma^{i}=\tau^{i}$. Thus, we choose the representation
\begin{align}
    \gamma^1 &= \begin{pmatrix}
       & 0 & & 1 & \\ & 1 & & 0 &
    \end{pmatrix}\,, & \gamma^2 &= \begin{pmatrix}
        & 0 & -i & \\ & i & 0 &
    \end{pmatrix}\,, & \gamma^3 &= \begin{pmatrix}
       & 1 & 0 & \\ & 0 & -1 &
    \end{pmatrix}\,.
\end{align}
Let us define the self-adjoint operator $K_{i}=i\Sigma_{i}$ and the ladder operator $K_{\pm}=K_{1}\pm iK_{2}$. These operators satisfy the Lie algebra of $SU(2)$, namely,
\begin{align}
    \left[K_i,K_j \right] &= i \epsilon_{ijk}K^k\,, & \left[K_3,K_\pm \right] &= \pm K_\pm\,, & \left[K_+,K_- \right] &= K_3\,.
\end{align}
For computing the meronic contribution to the spectrum, we also define $c_{(\pm)}=c_{(1)}\pm ic_{(2)}$. Denoting $\bar{K}_{x}=K_{x}-c_{(x)}$, where $x=\pm
,3$, as the shifted self-adjoint operator, the Dirac operator can be written as
\begin{align}
    \slashed{D} = \left[\begin{matrix}
        \alpha^{-1}\left(\bar{K}_3 + \frac{\alpha^2+2}{4}\right) & \bar{K}_-  \\ \bar{K}_+  & -\alpha^{-1}\left(\bar{K}_3 - \frac{\alpha^2+2}{4}\right)
    \end{matrix} \right]\,.
\end{align}
Following Ref.~\cite{HITCHIN19741,Bakas:2011nq}, we consider the direct sum of unitary irreducible
representation of $SU(2)$, say $|j,m\rangle$, with all $j=0,1/2,1,3/2,\ldots$
including half-integer values. The self-adjoint operators $K_{i}$ acting on
the state $|j,m\rangle$ give
\begin{align}
K_{\pm}|j,m\rangle &  =\sqrt{(j\mp m)(j\pm m+1)}\,|m\pm1\rangle\,,\\
K_{3}|j,m\rangle &  =m|j,m\rangle\,.
\end{align}
Thus, defining $p\equiv j+m+1$ and $q\equiv j-m$, the eigenvalues of the Dirac operator can be obtained by solving the characteristic polynomial $\det\left(\slashed{D}-\lambda\mathbb{I}\right)  =0$, giving
\begin{equation}
\lambda_{\pm}=\frac{\alpha}{4}\pm\frac{2}{\alpha}\sqrt{\left(  p-q-2c_{(3)}\right)^{2}+4\alpha^{2}\left(  c_{(1)}^{2}+c_{(2)}^{2}+pq\right)
-8\alpha^{2}c_{(1)}\sqrt{pq}}\,,\label{lambdapm}%
\end{equation}
for $q\neq0$. On the other hand, if $q=0$, the correct eigenvalue is obtained only for the upper sign of Eq.~\eqref{lambdapm} (see for instance Refs.~\cite{HITCHIN19741,Bakas:2011nq}), leading to
\begin{equation}
\lambda_{0}=\frac{\alpha}{4}+\frac{2}{\alpha}\sqrt{(p-2c_{(3)})^{2}%
+4\alpha^{2}\left(  c_{(1)}^{2}+c_{(2)}^{2}\right)  }\,.\label{lambda0}%
\end{equation}
The above expressions for $\lambda_{\pm}$ and $\lambda_{0}$ allow us to identify the non-Abelian contributions by comparing them with the analog expressions obtained in Refs.~\cite{Pope:1978zx,Pope:1981jx,Franchetti:2017ftp,Colipi-Marchant:2023awk} in the case of a squashed sphere with an Abelian field. In particular, if in Eqs.~\eqref{lambdapm} and~\eqref{lambda0} one takes $c_{(1)}=c_{(2)}=0$, then one obtains the eigenvalues for the Dirac operator derived in Refs.~\cite{HITCHIN19741,Bakas:2011nq}. Besides the intrinsic interest of these results, it opens the perspective to compute the parity anomaly~\cite{Redlich:1983kn,Redlich:1983dv,Alvarez-Gaume:1984zst,Korchemsky:1990va,Witten:2016cio,Kurkov:2017cdz,Kurkov:2018pjw} and the APS $\eta$-invariant~\cite{APS} identifying explicitly the non-Abelian contributions. We hope to come back on this important issue in a future publication.\footnote{In the Appendix, we have included some details that could be useful in the computation of the APS $\eta$-invariant, together with a discussion of the technical problems to be solved to achieve it.}

\section{Conclusions and perspectives\label{sec:conclusions}}

In this work, we have constructed the first analytic examples of regular self-gravitating anisotropic merons in the Einstein-Yang-Mills-Chern-Simons theory in three dimensions. The regular gauge field configurations have different meronic parameters along the three Maurer-Cartan forms for $SU(2)$. Moreover, the solution is topologically nontrivial as their Chern-Simons invariant is nonzero. Therefore, it cannot be continuously deformed into a trivial vacuum. The corresponding backreacted metric is conformally a squashed three-sphere which is regular as well. Therefore, these configurations represent gravitating non-Abelian solitons with nonvanishing topological charge. 

The amount of anisotropy of the gauge fields can be computed explicitly in terms of the squashing parameter, as well as the coupling constants of the theory. Indeed, we find that, in the large-squashing limit, the amount of anisotropy is finite; contrary to what happens when $\alpha\to0$. Moreover, the Dirac spectrum on this background for a spin-1/2 spinor in the fundamental representation of the gauge group can be computed explicitly and the genuine non-Abelian contributions to the spectrum can be identified. 

There are many relevant implications of the present construction. First of all, it opens the possibility to compute explicitly the APS $\eta$-invariant in a situation in which both genuine non-Abelian and gravitational contributions are present. Secondly, it clarifies---at least in the case of Einstein-Yang-Mills-Chern-Simons theory---what it means, precisely, to take the squashing parameter to be small or large; these two limits are often needed in holographic applications. Our results allow us to express explicitly the squashing parameter in terms of the coupling constant of the theory so that a large squashing can be obtained, for instance, for a large value of the Yang-Mills couplings. Consequently, a large squashing is equivalent to a strong-coupling expansion. Additionally, we found that small values for the gauge couplings favor less isotropic/symmetric gravitating solitons. Finally, analyzing the contribution of these gravitating anisotropic merons to the parity anomaly is certainly of great interest, since it has been shown that this phenomenon is closely related to magnetoresistance in three-dimensional metals~\cite{Goswami:2015uxa}, as well as topological responses and other chiral effects in condensed matter systems~\cite{Zyuzin:2012tv}. We will come back to this in the future.

\begin{acknowledgments}
The present authors would like to warmly thank Julio Oliva who participated in the early stages of this project, as well as Marcelo Oyarzo and Daniel Flores-Alfonso for illuminating discussions and suggestions. We also thank Francisco Colipí-Marchant, Marcela Lagos, Leonardo Sanhueza, Aldo Vera, and Jorge Zanelli for their insightful comments. The work of C.C. is partially supported by Agencia Nacional de Investigaci\'{o}n y Desarrollo (ANID) through FONDECYT grants No~11200025, 1230112, and~1210500. F.C. is supported by ANID through FONDECYT grant No~1200022. The Centro de Estudios Científicos (CECs) is funded by the Chilean Government through the Centers of Excellence Base Financing Program of Conicyt.
\end{acknowledgments}

\appendix

\section{On the APS eta-invariant\label{sec:eta}}

The main ingredient to compute the APS $\eta$-invariant is the explicit expression of the eigenvalues, which we have found explicitly in Eqs.~\eqref{lambdapm} and~\eqref{lambda0}. However, the issues related to the
zeta-function regularization become rather involved in the presence of the non-Abelian contributions. Here, we discuss why the technique of Ref.~\cite{Bakas:2011nq} cannot be applied directly; at first
glance, the latter is the most appropriate approach as far as the present case is concerned. 

The APS $\eta$-invariant is a measure of the spectral asymmetry of the Dirac operator~\cite{APS}. Recently, it has been reinterpreted as the axial charge of Dirac spinors~\cite{Kobayashi:2021jbn}. From the eigenvalues in Eqs.~\eqref{lambdapm}
and~\eqref{lambda0}, it can be computed by performing the analytic continuation of the meromorphic function $\eta_{D}(s)$ to $s=0$, i.e. \begin{align}
    \eta_D(s) = \sum_{\lambda\neq0} \sign(\lambda)\lambda^{-s}\,.
\end{align}
This term is invariant under the constant rescaling of the eigenvalues and it is divergent without the Riemann zeta-function regularization as it involves the sum of infinite terms. Each eigenvalue $\lambda_{\pm}$ has a
degeneracy of $2j+1$, while $\lambda_{0}$ has $2(2j+1)$; these values have to be taken into account when evaluating the $\eta$-invariant for the self-gravitating anisotropic meron. Then, $\eta_D(s)$ can be separated into three contributions; they are,
\begin{align}\label{etadecomp}
    \eta_D(s) = \sum_{p,q>0}(p+q)(\lambda_+^{-s} - \lambda_-^{-s}) + \sum_{p>0}2p\lambda_0^{-s} \equiv \eta_D^{(+)}(s) - \eta_D^{(-)}(s) + \eta_D^{(0)}(s) \,.
\end{align}

A natural approach to compute the $\eta$-invariant on squashed spheres is to replace the solution~\eqref{solk0} and perform $\lambda\rightarrow
\alpha\lambda$, using the fact that $\eta_D(s)$ remains invariant under constant rescalings of the eigenvalues. Then, one can perform a series expansion of the eigenvalues on the squashing parameter $\alpha$ as done in Refs.~\cite{HITCHIN19741,Bakas:2011nq}. This procedure yields
\begin{subequations}
\begin{align*}
\lambda_{+}^{-s} &  =\frac{(p-q-2)^{-s}}{2^{s}}\left[  1-\frac{2s\alpha
}{(p-q-2)}-\frac{s[16(pq-s-1)-11(p-q-2)]\alpha^{2}}{8(p-q-2)^{2}}%
+\ldots\right]  ,\\
\lambda_{-}^{-s} &  =\frac{(p-q-2)^{-s}}{(-2)^{s}}\left[  1-\frac{2s\alpha
}{(p-q-2)}-\frac{s[16(pq-s-1)-13(p-q-2)]\alpha^{2}}{8(p-q-2)^{2}}%
+\ldots\right]  ,\\
\lambda_{0}^{-s} &  =\frac{\left(  p-2\right)  ^{-s}}{2^{s}}\Bigg[1-\frac
{2s\alpha}{(p-2)}+\frac{s[16(s+1)+11(p-2)]\alpha^{2}}{8(p-2)^{2}}\nonumber\\
&  \qquad-\frac{s[128+64s(s+3)+(228+132s)(p-2)+27(p-2)^{2}]\alpha^{3}%
}{48(p-2)^{3}}+\ldots\Bigg]\,.
\end{align*}
Using the last expansion, the third contribution of Eq.~\eqref{etadecomp} can
be expressed in terms of the Riemann zeta function $\zeta(s)=\sum
_{n=1}^{\infty}n^{-s}$ as
\end{subequations}
\begin{align}
\eta_{D}^{(0)}(s) &  =\frac{1}{(-2)^{s}}\left\{  2+4s\alpha+\frac
{s(16s+5)\alpha^{2}}{4}+\frac{s\left(  64s^{2}+60s-73\right)  \alpha^{3}}%
{24}\right\}  \nonumber\\
&  +\frac{1}{2^{s-1}}\Bigg\{\zeta(s-1)+\left[  2-2s\alpha+\frac{11s\alpha^{2}%
}{8}-\frac{27s\alpha^{3}}{48}\right]  \zeta(s)\nonumber\\
&  -\bigg[4\alpha-\frac{1}{4}\left(  19+8s\right)  \alpha^{2}+\frac{1}%
{8}\left(  47+22s\right)  \alpha^{3}\bigg]s\zeta(s+1)+\ldots\Bigg\}\,.
\end{align}
We have omitted all the terms $s\zeta(s+n)$ for $n\geq2$ since they vanish in
the limit $s\rightarrow0$. Additionally, we know that $\zeta(s)$ is a
meromorphic function in the whole complex plane that has a simple pole in
$s=1$ with residue $1$, which implies that $s\zeta(s+1)=1$ as $s\rightarrow0$.
Then, to quintic order in the squashing parameter, we find that the contribution of $\lambda_{0}$ to the $\eta$-invariant is
\begin{align}
\eta_{D}^{(0)}=\lim_{s\rightarrow0}\eta_{D}^{(0)}(s)=-\frac{1}{6}%
-8\alpha+\frac{19}{2}\alpha^{2}-\frac{47}{4}\alpha^{3}+\frac{457}{64}%
\alpha^{4}-\frac{603}{256}\alpha^{5}+\mathcal{O}(\alpha^{6})\,.    
\end{align}

The contributions of $\lambda_{\pm}^{-s}$ to the $\eta$-invariant are more
involved. First, we notice that the presence of the non-Abelian gauge fields to the eigenvalues $\lambda_\pm$ introduce nonlinear dependence on the squashing parameter that renders its expansion more complicated than in the case of $\lambda_0$. This implies that the form of the meromorphic function found in Refs.~\cite{HITCHIN19741,Bakas:2011nq} cannot be generalized directly to this case. Therefore, a thorough analysis of their poles and residues' structure has to be done carefully before computing the remaining contributions of the $\eta$-invariant. We postpone a deeper analysis of this problem for a forthcoming paper since new techniques are required to compute the APS $\eta$-invariant in this case. 

\bibliography{References}

\end{document}